\documentclass[journal,twoside]{IEEEtran}

\usepackage{amsmath}
\usepackage{amssymb}
\usepackage[dvips]{graphicx}
\graphicspath{{./fig/}}
\usepackage{curves,color}
\usepackage{subfigure}
\usepackage{dsfont,slashbox,wasysym}
\usepackage{tikz}
\usepackage{pdfpages,enumerate}
\usepackage{multirow}
\usepackage{bm}
\usepackage{algorithmic}
\usepackage{algorithm}

\allowdisplaybreaks

\RequirePackage[normalem]{ulem} 
\RequirePackage{color}\definecolor{RED}{rgb}{1,0,0}\definecolor{BLUE}{rgb}{0,0,1} 

\begin{document}
\title{Iterative Soft Decoding Algorithm for DNA Storage Using Quality Score and Redecoding}
\author{Jaeho Jeong, \IEEEmembership{Graduate Student Member, IEEE}, Hosung Park, \IEEEmembership{Member, IEEE}, Hee-Youl Kwak, \IEEEmembership{Member, IEEE}, Jong-Seon No, \IEEEmembership{Fellow, IEEE}, Hahyeon Jeon, Jeong Wook Lee, and Jae-Won Kim, \IEEEmembership{Member, IEEE}
\thanks{Manuscript received xxxx xx, 2022; revised xxxx xx, 20xx; accepted xxxx xx, 20xx. Date of publication xxxx xx, 20xx; date of current version xxxx xx, 20xx. This work was supported in part by Samsung Research Funding and Incubation Center of Samsung Electronics under Project Number SRFC-IT1802-09 and in part by the Pioneer Research Center Program through the National Research Foundation of Korea funded by the Ministry of Science, ICT \& Future Planning (No. \textcolor{black}{NRF-2022M3C1A3081366}). \textit{(Corresponding author: Jae-Won Kim.)}}
\thanks{Jaeho Jeong and Jong-Seon No are with Department of Electrical and Computer Engineering, Institute of New Media and Communications (INMC), Seoul National University, Seoul 08826, South Korea (e-mail: jaehoj@ccl.snu.ac.kr; jsno@snu.ac.kr).}
\thanks{Hosung Park is with Department of Computer Engineering and Department of ICT Convergence System Engineering, Chonnam National University, Gwangju 61186, South Korea (e-mail: hpark1@jnu.ac.kr).}
\thanks{Hee-Youl Kwak is with School of Electrical Engineering, University of Ulsan, Ulsan 44610, South Korea (e-mail: hykwak@ulsan.ac.kr).}
\thanks{Hahyeon Jeon is with Clinomics, Ulsan 44919, South Korea (e-mail: jeonhh0061@gmail.com).}
\thanks{Jeong Wook Lee is with Department of Chemical Engineering, POSTECH, Pohang 37673, South Korea (e-mail: jeongwook@postech.ac.kr).}
\thanks{Jae-Won Kim is with Department of Electronic Engineering, Engineering Research Institute (ERI), Gyeongsang National University, Jinju 52828, South Korea (e-mail: jaewon07.kim@gnu.ac.kr).}
\thanks{Digital Object Identifier xxxxxxxxxxxx}
}

\maketitle

\begin{abstract}
Ever since deoxyribonucleic acid (DNA) was considered as a next-generation data-storage medium, lots of research efforts have been made to correct errors occurred during the synthesis, storage, and sequencing processes using error correcting codes (ECCs). Previous works on recovering the data from the sequenced DNA pool with errors have utilized hard decoding algorithms based on a majority decision rule. To improve the correction capability of ECCs and robustness of the DNA storage system, we propose a new iterative soft decoding algorithm, where soft information is obtained from FASTQ files and channel statistics. In particular, we propose a new formula for log-likelihood ratio (LLR) calculation using quality scores (Q-scores) and a redecoding method which may be suitable for the error correction and detection in the DNA sequencing area. Based on the widely adopted encoding scheme of the fountain code structure proposed by Erlich {\em et al.}, we use \textcolor{black}{three} different sets of sequenced data to show consistency for the performance evaluation. The proposed soft decoding algorithm gives 2.3\%$\sim$7.0\% improvement of the reading \textcolor{black}{number} reduction compared to the state-of-the-art decoding method and it is shown that it can deal with erroneous sequenced oligo reads with insertion and deletion errors.
\end{abstract}


\vspace{2pt}
\section{Introduction}
\label{sec:Introduction}
\IEEEPARstart{A}{s huge} amounts of data are rapidly produced in the era of big data, there is a need for a new type of high-density storage media that is more competitive than existing ones such as magnetic tapes or hard disk drives. Major information technology companies are spending enormous financial resources for building new data centers to cope with the increased storage demand. With the development of new synthesis and sequencing technologies, deoxyribonucleic acid (DNA) has been considered as a competitive candidate for new storage media \cite{Church}. DNA is known to have up to more than hundreds of million times information density compared to the existing storage media and can survive thousands of years when the temperature is kept appropriately \cite{Dong}. Since the pioneering study in \cite{Church}, the idea of using DNA as a storage medium has attracted a significant attention and a lot of studies are going on to make use of the advantages of its longevity and high information density nowadays.

As the data storage device, we consider single-stranded DNA consisting of four nucleotides (i.e., bases). These four bases are Adenine, Cytosine, Guanine, and Thymine, which are simply represented as A, C, G, and T, respectively. We refer to this single-stranded DNA as an oligonucleotide or simply an oligo. For the DNA storage system, writing and reading processes are called DNA synthesis and sequencing, respectively. In order to read information stored in oligos through sequencing, polymerase chain reaction (PCR) that amplifies the information is needed in advance. In fact, each oligo is synthesized in large numbers and the synthesized number for each oligo is not uniform. Furthermore, each piece of information stored in the oligo can be amplified to a different extent during PCR amplification. Using the sequencing equipment, sequenced data is usually produced in computer file formats. Among them, FASTQ files had become one of the common formats in the DNA sequencing area \cite{Cock} by providing a quality score (Q-score), which is a probability of how reliably each base is predicted and sequenced. In all steps described above, three types of errors can occur, that is, substitution errors, insertion errors, and deletion errors. The error distributions and behaviors are very different from those of the conventional memory devices. While substitution errors have mainly been considered in the conventional memory devices, the other types of errors are rarely considered. To handle these errors, it is necessary to adopt carefully designed error correcting codes (ECCs) for data.

To guarantee data reliability for DNA storage, there are some biochemical constraints on data such as the oligo length, maximum homopolymer-run length, and GC-content \cite{Xu}, \cite{El-Shaikh}. The \textcolor{black}{lengths} of oligos are typically from hundreds to a few thousands and it is reported that longer oligos are hard to be synthesized and prone to errors during several steps. In \cite{Ross} and \cite{Schwartz}, the effects of the maximum homopolymer-run length and GC-content were discussed. It was shown that long homopolymer-run length of oligos increases error rates and a moderate GC-content is needed for PCR amplification to avoid high dropout rates. Also, studies in \cite{Cao_nano}--\cite{Yin} suggested that some constraints such as the minimum free energy (MFE) constraint, end-constraint, self-complementary constraint, distances between oligo sequences, and correlation between addresses should be considered to reduce the error rate of DNA storage.

In addition, there are many works on new coding schemes for the DNA storage system \cite{Goldman}--\cite{Jeong}. Organick {\em et al.} \cite{RandomAccess} designed a large number of primers and validated that the individual recover of files is possible. In the study of Yazdi {\em et al.} \cite{Portable}, an iterative alignment algorithm was introduced and they converted all types of errors into deletion errors, while Erlich {\em et al.} \cite{Science} achieved 86\% of the theoretical limit (i.e., capacity) through the encoding scheme using fountain codes and Reed-Solomon (RS) codes. Also, Chandak {\em et al.} \cite{Chandak} developed a single large block code structure using low-density parity-check (LDPC) codes with soft decoding using log-likelihood ratio (LLR)\textcolor{black}{.} Press {\em et al.} \cite{Hedges} handled insertion and deletion errors using a convolutional code with a hash function. Additionally, Wang {\em et al.} \cite{Wang} designed an oligo structure using only one primer binding site (PBS) for \textcolor{black}{higher} information density and Cao {\em et al.} \cite{Cao_npj} applied \textcolor{black}{adaptive} coding using fountain codes to satisfy several constraints mentioned above \cite{Ross}--\cite{Yin}. In the end, our previous work \cite{Jeong} achieved a decoding performance improvement by combining several preprocessing components and decoding techniques while the encoding scheme is the same as \textcolor{black}{that of \cite{Science}}.

In this paper, we propose an iterative soft decoding algorithm based on redecoding using RS codes. We follow the encoding scheme of \cite{Science} like our previous works \cite{Jeong}, \cite{Kang} and show the performance improvement by modifying the decoding algorithm. In particular, we decode the received codewords (i.e., sequencing results) using soft information provided by FASTQ files, that is, Q-scores for oligo reads. In the previous work \cite{Science}, they used RS codes as error detecting codes to check whether sequenced reads of oligos are erroneous. If the sequenced oligo reads are erroneous, those oligo reads are discarded, which means that the decoding method in \cite{Science} requires a larger number of sequenced oligo reads.

\textcolor{black}{From the decoding performance of the proposed algorithm, we show} that the soft information acquired from FASTQ files can reduce the required number of sequenced oligo reads and accordingly save the \textcolor{black}{reading cost} of the sequencing process. While the LLR calculation in \cite{Chandak} for soft decoding was based on the hard values (i.e. the difference \textcolor{black}{between} the \textcolor{black}{numbers} of 0s and 1s), we determine the LLR values based on the probability of the correct basecall from the Q-scores and channel statistics for \textcolor{black}{entire} oligo reads. Since FASTQ files only offer a Q-score for one nucleotide in each position, we estimate the probabilities for the other nucleotides with the channel statistics considering edit distances. From the basecalling probabilities of the four bases in each position of oligo reads, we can obtain an accurate estimate of the LLR values and perform soft decoding with high performance.

\begin{figure*}[t]
    \centering
    \includegraphics[scale=0.25]{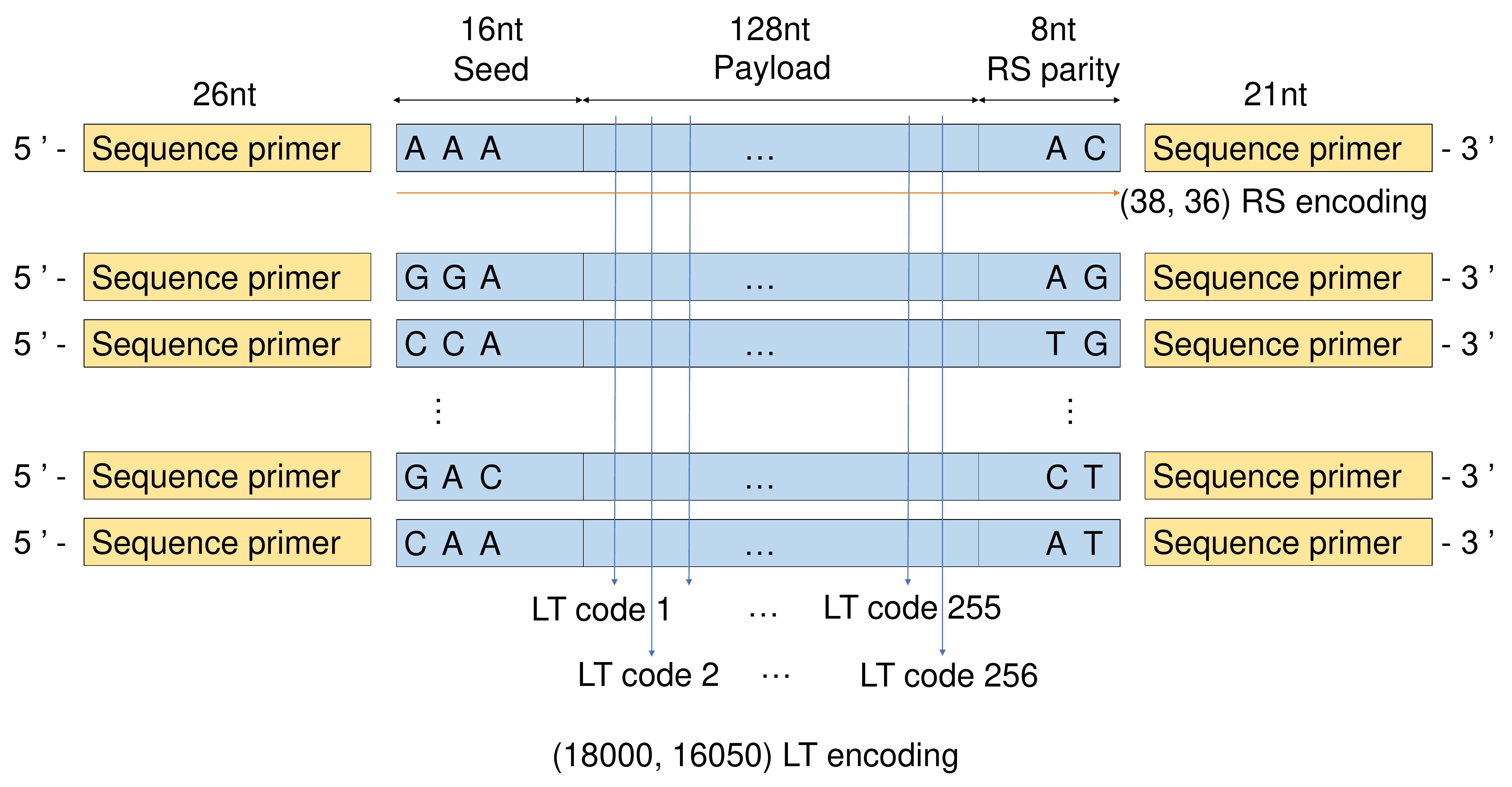}
    \caption{\textcolor{black}{The encoding structure of our experiment. We use the (18000, 16050) LT code for the inter-oligo code and the (38,36) RS code for the intra-oligo code, in which four nucleotides make one RS symbol.}}
    \label{fig1}
\end{figure*}

Compared to our previous work \cite{Jeong} in which the main decoding process stays unchanged from \cite{Science}, this work presents a novel decoding process summarized as follows:
\begin{itemize}
  \item We first propose a new formula for LLR caluclation utilizing soft information from FASTQ files, which is the first attempt in the DNA storage field and can inspire diverse studies in the DNA sequencing area.
  \item We introduce a new iterative soft decoding algorithm using the redecoding method based on RS decoding results. It is shown that this redecoding method can extract and discard erroneous sequenced oligo reads with insertion and deletion errors.
  \item We use \textcolor{black}{three} different sets of sequenced data to show the performance improvement of the proposed decoding algorithm compared to the state-of-the-art algorithm \textcolor{black}{\cite{Jeong}}.
\end{itemize}

This paper is organized as follows. In \textcolor{black}{Section~\ref{sec:MaterialsandMethods}}, we cover our experimental design including the synthesis and sequencing information. We also explain the proposed iterative soft decoding algorithm and how each technique works. In \textcolor{black}{Section~\ref{sec:Result}}, we show the superiority of the proposed methods compared to existing decoding methods and give analysis on the experimental results. In \textcolor{black}{Section~\ref{sec:DiscussionandFutureWork}}, we discuss several considerations that we could not include in this paper and expects for the future works. Finally, conclusion is given in \textcolor{black}{Section~\ref{sec:Conclusion}}.

\section{Materials and Methods}
\label{sec:MaterialsandMethods}
Most of the previous works in DNA storage rely on hard decoding for the decoding process after the sequencing process is done. However, research in \cite{Chandak} showed that soft decoding can be applied for the DNA storage system. Even though we employ a different code structure from \cite{Chandak}, we adopt LLR calculation and the soft decoding into our experiment.

\subsection{Experimental Design}
\label{subsec:ExperimentalDesign}
We \textcolor{black}{first} use two different sequenced data sets to show the flexibility and consistency of our proposed methods. One sequenced data set is the experimental data used for our previous work \cite{Jeong} and we call it $data A$. For the other set, called $data B$, we conduct another sequencing experiment from the remaining synthesized pool in \cite{Jeong} with the same environmental conditions.

\textcolor{black}{Our encoding structure is shown in Fig.~\ref{fig1}.} As we mentioned in \cite{Jeong}, we use the Luby transform (LT) code \cite{Luby} for the inter-oligo code and the RS code for the intra-oligo code. The length of each oligo is 152 nucleotides (nt)\textcolor{black}{, where each oligo is composed of 16nt, 128nt, and 8nt for the seed, payload, and RS parity, respectively. For the RS code, a parity symbol is 4nt each and we use the (38, 36) RS code in the} \textcolor{black}{finite field of $GF(2^8)$, which makes a minimum Hamming distance be 3, implying 1-symbol correction or 2-symbol detection is available.} Using LT encoding with the robust-soliton distribution \cite{Science}, we make 18000 oligo sequences from 16050 information sequences \textcolor{black}{converted from the image file of 513.6 KB. From the mapping between the binary data and DNA bases, we transform two consecutive bits into a base, that is, 00 = A, 01 = C, 10 = G, and 11 = T. We add two sequence primers at the ends of each oligo sequence:}
\begin{itemize}
\color{black}
  \item 5' side: GTTCAGAGTTCTACAGTCCGACGATC,
  \item 3' side: TGGAATTCTCGGGTGCCAAGG.
\end{itemize}

In \cite{Jeong}, two oligo pools were synthesized to show the effects of homopolymer-run length and GC-content constraints. Among them, we \textcolor{black}{first} consider the non-constrained pool \textcolor{black}{($data A$ and $data B$)} in this \textcolor{black}{section}, which allows us to have the seed information of the total oligo pool during the decoding process. The non-constrained pool means that we do not contemplate any homopolymer-run length or GC-content constraints for encoding. Considering the encoding structure for the non-constrained pool \cite{Jeong}, it is easily noticed that we just directly create 18000 different seed structures of LT codes on the encoding process. It means that we use the first 18000 seeds of the robust-soliton distribution \cite{Science} and we can get exactly the same seed structures when we follow the first 18000 seed calculation at the decoding process. Since the purpose of this paper is to improve the decoding performance, we try to restrict other conditions such as exactness of seed information and several preprocessing steps. For this reason, the non-constrained pool is more suitable for comparison between different decoding algorithms. Further, the decoding performance of the non-constrained pool was comparable to that of the constrained pool in the previous study \cite{Jeong}, which means that errors induced by not satisfying the homopolymer-run length and GC-content constraints can be reasonably covered by decoding algorithms.

For the new experiment with $data B$, we use the same sequencing process \textcolor{black}{applied} for the non-constrained pool in \cite{Jeong}. The ratio of the GC-content \textcolor{black}{is} $[0.3289, 0.6842]$ and the maximum homopolymer-run length is 13. We use RP1 and RPI1 primers for PCR amplification and Illumina Miseq Reagent v3 kit for the sequencing with 600 cycles, reading 151nt each in forward and reverse directions. Q30 of this new sequencing experiment is 96\% and output contains 22 million forward-reads and 22 million reverse-reads, in total 44 million sequenced oligo reads, leaving 36\% PhiX spike-in.

\subsection{Proposed Decoding Algorithm}
\label{subsec:ProposedDecodingAlgorithm}
\begin{figure}[!t]
    \includegraphics[scale=0.42]{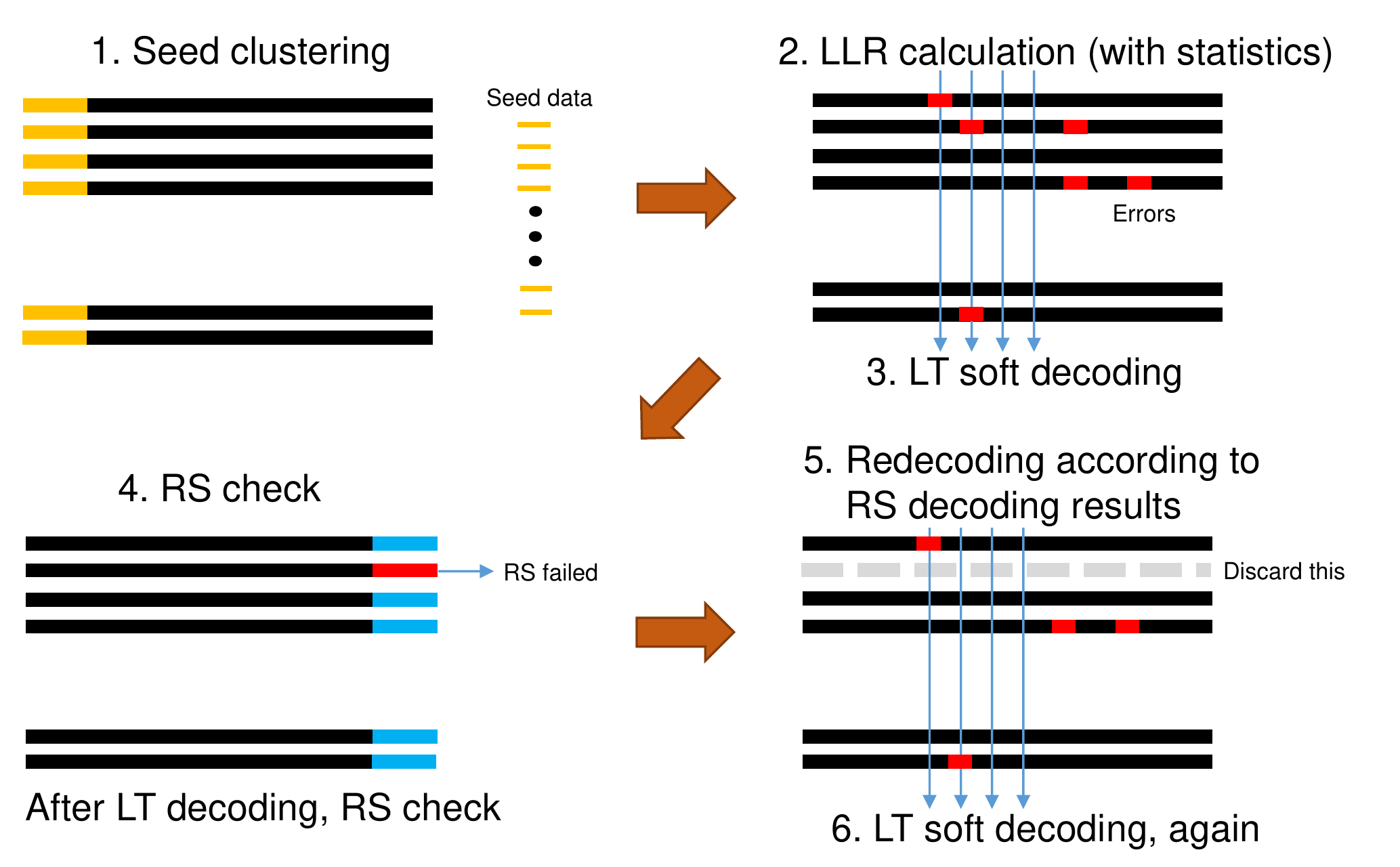}
    \caption{The description for the proposed decoding process.}
    \label{fig2}
\end{figure}

\begin{figure}[!t]
    \includegraphics[scale=0.31]{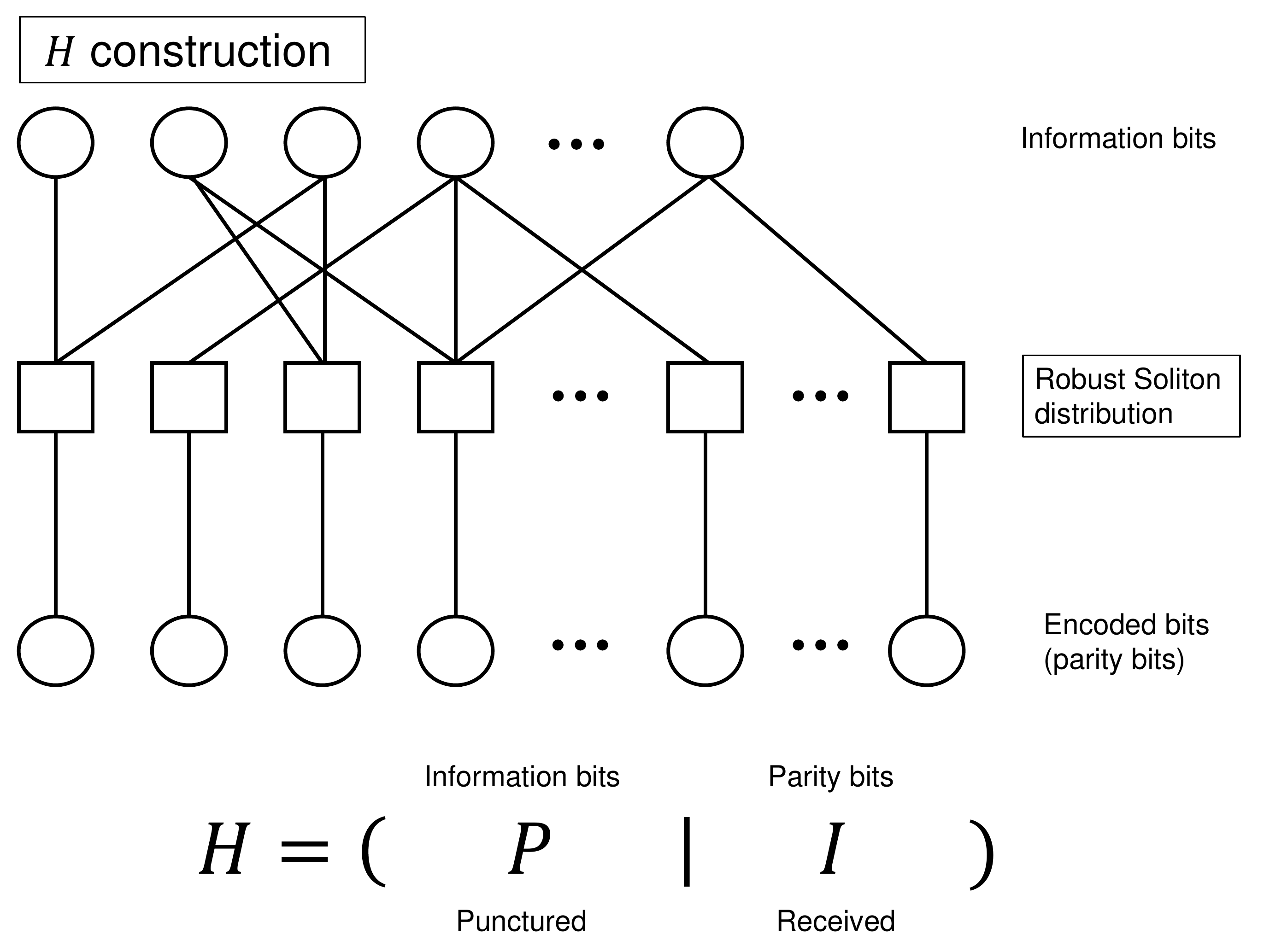}
    \caption{The construction of a parity check matrix $H$.}
    \label{fig3}
\end{figure}

The proposed decoding process is shown in Fig.~\ref{fig2}\textcolor{black}{, where each process will be explained in the following subsections}. In the previous work \cite{Jeong}, we used the oligo clustering that binds similar sequenced oligo reads into a cluster by comparing the Hamming distance of the whole oligo sequences. In this work, we use seed clustering that binds sequenced oligo reads with exactly the same seed structure into a cluster by comparing only the seed part of the sequenced oligo reads. Since we have the seed information of the oligo pool by calculating the robust-soliton distribution, we can decide that which seed structure is erroneous and which seed structure is correct \textcolor{black}{(although undetected errors can exist)}. Then, we only utilize the clusters with the \textcolor{black}{pre-determined} seed structure and discard clusters that have \textcolor{black}{different seed parts}. Using this seed information, we make a parity check matrix $H$ for LT soft decoding as shown in Fig.~\ref{fig3}. \textcolor{black}{In order to construct the valid parity check matrix, the decoder has to know the pre-determined seed values and it is why we consider the non-constrained pool.}

For LT codes, the number of symbols $K$ required for LT decoding with a failure rate $\delta$ is derived in \textbf{Theorem 12} of \cite{Luby} as follows:
\begin{equation}
K = k\beta = k (\sum_{i}^{}\rho(i)+\tau(i)) = k + \sum_{i=1}^{k/R-1}\frac{R}{i}+R\ln(R/\delta),
\nonumber
\end{equation}
where $k$ equals to the number of information packets and $R=c\times {\rm ln}({k \over \delta})\times \sqrt {k}$ for a constant value $c > 0$. By setting parameters capturing our experimental conditions, we could have $K = 16951$ with $\delta = 0.001$ and $c = 0.025$. Based on the calculated value $K$, we need to make sure that the number of clusters participating for LT decoding should be at least $K$ or more during the decoding process. In Fig.~\ref{fig3}, the number of information bits equals to 16050 and the number of parity bits should be in a range $16951 \leq K \leq 18000$. After constructing $H$ matrix using seed information, we perform belief propagation (BP) decoding \cite{Han} with this $H$ matrix for 500 decoding iterations.

Our DNA oligo pool has a product structure of the LT code and RS code, and thus we can choose which code to be decoded first. The previous work \cite{Jeong} decoded the RS code first and LT code later following the similar decoding process in \cite{Science}. At that work, we used the RS code for the error correction and if the decoded codeword had errors in it, this cluster would give a terrible effect over the whole LT decoding process. This is because the LT code only worked as a prevention for a loss of each oligo sequence while not giving any help for the error correction. In this work, however, we decode the LT code first and RS code later. We use soft decoding for the LT code and use it for the error correction and use the RS code for both the error correction and error detection. In this scenario, if there are errors in the LT correction, we can find them using the RS code. Specifically, we utilize the RS code for two purposes: i) error detection for seed positions and ii) error correction for other positions. If the RS decoding fails or error is detected for the seed positions, we discard that erroneous information and repeat the LT soft decoding again by constructing a new $H$ matrix while removing the erroneous oligo reads. In the following subsections, we explain these decoding processes in detail.

\subsubsection{LLR Calculation}
\label{subsubsec:LLRCalculation}
Since the inputs of BP decoding are LLR values for each bit, we need to extract LLR values from the soft information \cite{Palanki}. In the previous work using soft decoding \cite{Chandak}, they used hard values for calculating LLR values based on basecalling. By mapping the basecalled hard values into 0s and 1s, they counted the \textcolor{black}{numbers} of 0s and 1s in a cluster for each position, and calculated LLR values according to the \textcolor{black}{numbers} of 0s and 1s with the heuristically determined crossover probability. On the other hand, we try to extract soft values provided in the sequenced FASTQ files, that is, Q-scores.

\begin{figure*}[t]
    \centering
    \includegraphics[scale=0.22]{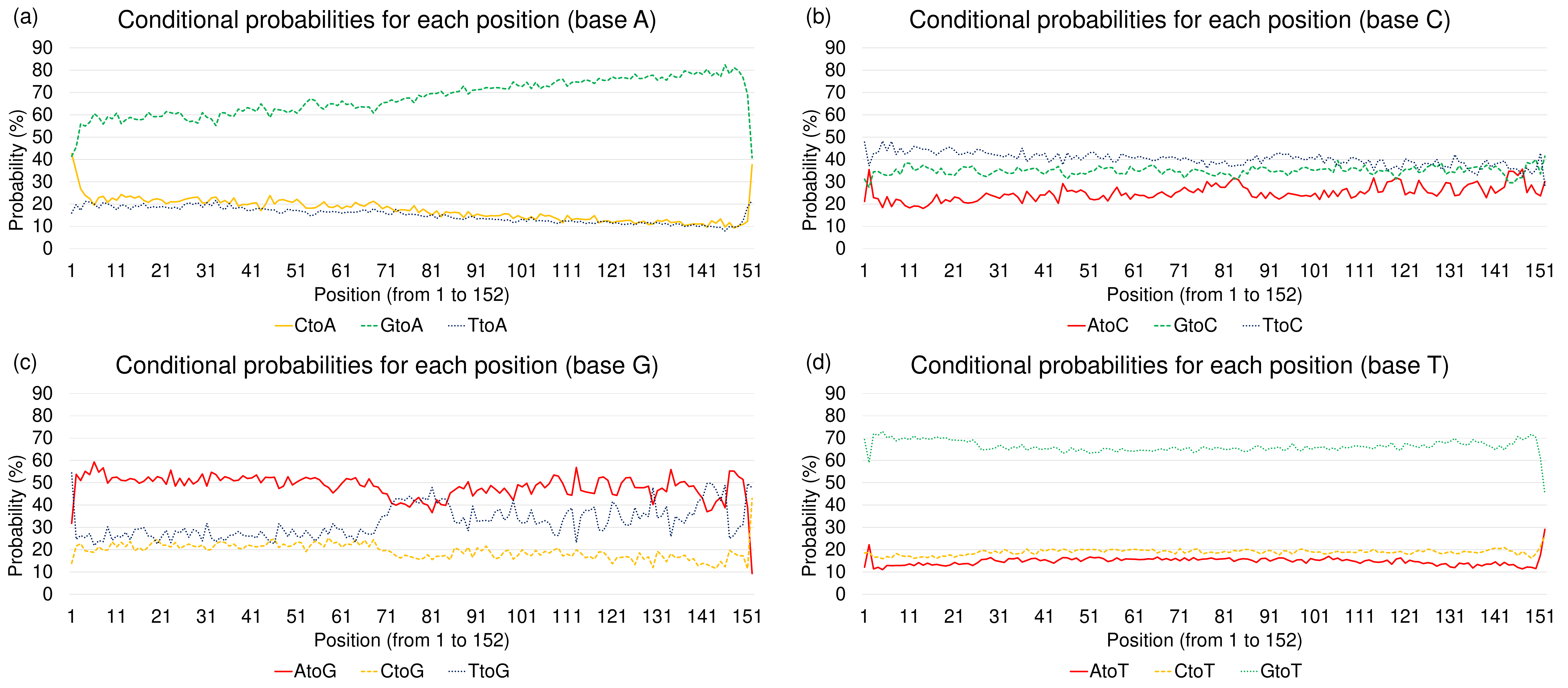}
    \caption{The \textcolor{black}{graphs} of \textcolor{black}{conditional probabilities of base transitions in position errors} for each position in $data B$. XtoY in the \textcolor{black}{graphs} means \textcolor{black}{a} probability that X is stored but Y is sequenced instead \textcolor{black}{(X $\neq$ Y)}. The probabilities are given from position 1 to 152. (a) \textcolor{black}{Conditional probabilites} for base A. (b) \textcolor{black}{Conditional probabilites} for base C. (c) \textcolor{black}{Conditional probabilites} for base G. (d) \textcolor{black}{Conditional probabilites} for base T. Note that the sums of \textcolor{black}{conditional} probabilities of \{CtoA, GtoA, TtoA\}, \{AtoC, GtoC, TtoC\}, \{AtoG, CtoG, TtoG\}, \textcolor{black}{and} \{AtoT, CtoT, GtoT\} all become 100\% \textcolor{black}{because they mean conditional base transition probabilities when an error occurs for each position.}}
    \label{fig4}
\end{figure*}

To obtain accurate LLR values, we use the error distribution statistics of the entire sequenced data from the oligo pool. We collect all the correct-length \textcolor{black}{(152nt)} stitched oligo reads using the PEAR algorithm \cite{Zhang} and compare them with the original encoded data to figure out the occurrence of errors in each sequenced oligo read. Then, for each position, we calculate \textcolor{black}{conditional rates} for every combination of base transition, i.e., A$\rightarrow$C, A$\rightarrow$G, A$\rightarrow$T, C$\rightarrow$A, C$\rightarrow$G, C$\rightarrow$T, G$\rightarrow$A, G$\rightarrow$C, G$\rightarrow$T, T$\rightarrow$A, T$\rightarrow$C, and T$\rightarrow$G, where X$\rightarrow$Y denotes the event that X is stored but Y is sequenced instead \textcolor{black}{(X $\neq$ Y)}. \textcolor{black}{Since these probabilities are conditional (for errors occuring), we calculate the conditional probabilities only in the cases of sequenced oligo reads with any error occurrences.} For the decoding process, we only consider position errors, which means that we handle the insertion and deletion errors by considering them as burst substitution errors. This is reasonable because the insertion and deletion error rates are very low compared to the substitution error rate when collecting only the correct-length stitched oligo reads. Furthermore, there is no way to correct insertion and deletion errors in the perspective of ECCs because we employ the encoding scheme in \cite{Science}. For this reason, we only consider and utilize correct-length (152nt) stitched oligo reads for decoding. By arranging the error distributions for each position, one can notice that there are some tendencies of transition probabilities (Fig.~\ref{fig4}) and this result motivates us to use \textcolor{black}{conditional probabilities of base transitions in position errors} from entire sequenced data for LLR calculation.

Now, we explain how to derive conditional \textcolor{black}{base transition} probabilities from entire sequenced data. Let $Z[n]=\{1,2,\ldots,n\}$ for a positive integer $n$ and $N_{\rm read}$ be the number of stitched oligo reads after processing the PEAR algorithm for entire sequenced data. For each stitched oligo index $j\in Z[N_{\rm read}]$, $g(j)$ denotes an index of the nearest encoded oligo sequence based on the edit distance. For each position index $i\in Z[152]$, we define $f_i(b^{(1)}\rightarrow b^{(2)})=\Sigma_j I(x^{i,g(j)}=b^{(1)}, y^{i,j}=b^{(2)})$, where $x^{m,n}$ is a base of the $m$th position in the $n$th encoded oligo sequence, $y^{m,n^\prime}$ is a base call of the $m$th position in the $n^\prime$th stitched oligo read for $m\in Z[152]$, $n\in Z[18000]$, and $n^\prime\in Z[N_{\rm read}]$, $I(\cdot)$ denotes an indicator function, and $b^{(1)},b^{(2)} \in \{{\rm A,C,G,T}\}$ such that $b^{(1)}\neq b^{(2)}$.

Let $h(j)$ denote an index of the real encoded oligo index corresponding to the $j$th stitched oligo read. Then, we define the $i$th conditional \textcolor{black}{base transition rates} as follow:
\begin{equation}
\begin{split}
& P_i (x^{i,h(j)}=b^{(1)}|y^{i,j}=b^{(2)})\\
& ={{f_i(b^{(1)}\rightarrow b^{(2)})\over{N_i(b^{(1)})}}\over{{f_i(b^{(1)}\rightarrow b^{(2)})\over N_i(b^{(1)})} + {f_i(b^{(3)}\rightarrow b^{(2)})\over N_i(b^{(3)})} + {{f_i(b^{(4)}\rightarrow b^{(2)})}\over N_i(b^{(4)})}}},
\label{eq:positionerror}
\nonumber
\end{split}
\end{equation}
where four distinct $b^{(1)},b^{(2)},b^{(3)}$, $b^{(4)}\in \{{\rm A,C,G,T}\}$ and $N_i(b^{(k)})=\Sigma_j I(x^{i,g(j)}=b^{(k)})$, that is, the number of $b^{(k)}$s in the $i$th position of encoded sequences corresponding to the stitched oligo reads for $k\in Z[4]$. The \textcolor{black}{conditional base transition probabilities} for entire sequenced data are given in Fig.~\ref{fig4}.

Now, we explain how to calculate LLRs using Q-scores and channel statistics. After classifying stitched oligo reads into clusters based on seed information, we calculate LLR values for each cluster. Let $y_1^{i,j}$ and $y_2^{i,j}$ denote the $1$st and $2$nd bits of the $i$th position of the $j$th stitched oligo read for a mapping $\{{\rm A,C,G,T}\}\rightarrow\{00,01,10,11\}$, $i\in Z[152]$, and $j\in Z[N_{\rm read}]$. Let $P_i^{j,b}(b^\prime)=P(x^{i,h(j)}=b^\prime|y^{i,j}=b)$ for $b,b^\prime\in\{{\rm A,C,G,T}\}$ and we determine this probability for the case with $b=b^\prime$ from the FASTQ file. The probability is derived as follow~\cite{Illumina}:
\begin{equation}
P_i^{j,b}(b) = 1-10^{- \frac{Q_i^j(b)}{10}},
\nonumber
\end{equation}
where $Q_i^j(b)$ is a Q-score value of the $i$th position of the $j$th stitched oligo read (and the basecall is $b$) in the FASTQ file. Although Q-scores do not represent all types of errors in DNA storage, it is natural to believe the higher Q-score value represents the higher reliability.

Then, we can calculate the other three probabilities of bases for a base call $b$ in that position as follow:
\begin{equation}
\begin{split}
& P_i^{j,b}(b^{(k)})=(1-P_i^{j,b}(b))\times P_i (x^{i,h(j)}=b^{(k)} | y^{i,j}=b),
\end{split}
\nonumber
\end{equation}
where $k\in Z[3]$ and $b^{(k)}\in\{{\rm A,C,G,T}\}$ such that $b^{(k)}\neq b$.

For example, suppose that base ${\rm A}$ is sequenced in the $s$th position of the $t$th stitched oligo read with a Q-score of 10 in the FASTQ file. Further, assume that we have error statistics of the $s$th position such that the conditional probabilities are $P_s(x^{s,h(t)}={\rm C}|y^{s,t}={\rm A})=50\%, P_s(x^{s,h(t)}={\rm G}|y^{s,t}={\rm A})=25\%, P_s(x^{s,h(t)}={\rm T}|y^{s,t}={\rm A})=25\%$. Then\textcolor{black}{,} we can obtain the probabilities as follows:
\begin{flalign}
& P_s^{t,{\rm A}}({\rm A}) = 1-10^{- \frac{10}{10}} = 0.9\nonumber,\\
& P_s^{t,{\rm A}}({\rm C}) = (1-P_s^{t,{\rm A}}({\rm A}))\times 0.5 = 0.05\nonumber,\\
& P_s^{t,{\rm A}}({\rm G}) = (1-P_s^{t,{\rm A}}({\rm A}))\times 0.25 = 0.025\nonumber,\\
& P_s^{t,{\rm A}}({\rm T}) = (1-P_s^{t,{\rm A}}({\rm A}))\times 0.25 = 0.025\nonumber.\\
\nonumber
\end{flalign}

After all, we can calculate LLRs of bits for a base call $b$ considering a mapping $\{{\rm A,C,G,T}\}\rightarrow\{00,01,10,11\}$ as follows:
\begin{flalign}
{\rm LLR}(y_1^{i,j})={\rm log}{{P_i^{j,b}({\rm A})+P_i^{j,b}({\rm C})}\over{P_i^{j,b}({\rm G})+P_i^{j,b}({\rm T})}}\nonumber,\\
{\rm LLR}(y_2^{i,j})={\rm log}{{P_i^{j,b}({\rm A})+P_i^{j,b}({\rm G})}\over{P_i^{j,b}({\rm C})+P_i^{j,b}({\rm T})}},
\nonumber
\end{flalign}
which are from the previous work \cite{Lu}.

\textcolor{black}{Based on the pre-determined} seed information, we can discard stitched oligo reads with \textcolor{black}{incorrect} seed information (basecall) for LLR calculation and decoding. For oligo reads with the same seed information, we simply add LLRs of them for LT soft decoding. Since soft decoding is used only for the LT code and seed positions utilize the basecall information, LLR calculation is only needed for 128 payload positions. For RS parity, we exploit Q-scores and it will be explained in \textcolor{black}{Section~\ref{subsubsec:IterativeDecoding}}.

\subsubsection{Redecoding Based on RS Decoding Results}
\label{subsubsec:RedecodingBasedonRSDecodingResults}
As aforementioned, we first proceed LT soft decoding and then we use the RS code for both error correction and error detection. After LT soft decoding using the parity check matrix $H$ constructed from the seed information, we perform RS decoding for the decoded oligos. If RS decoding succeeds for all of the decoded oligos without any seed position errors, we recover the data from the doubly decoded oligos and declare an overall decoding success or failure. However, most of the cases do not satisfy the above conditions at once. Here, we propose the redecoding method that performs redecoding based on RS decoding results. For LT decoded oligos with RS decoding failures or any error \textcolor{black}{detection} in the seed positions, we discard the corresponding oligo clusters and then we proceed LT soft decoding again. This means that we do not use these wrong-RS-checked clusters into the decoding process, making a new $H$ matrix with the same LLR calculation for the other oligo reads. Further, it is natural to regard the RS corrected oligos with seed position errors as wrongly corrected oligos because we only utilize oligo reads with the \textcolor{black}{pre-determined seed value. Later, we actually find out that there are still some RS-corrected oligo clusters with the correction on the seed part, and they get removed accurately during our experiment.}

In fact, this redecoding operation helps us to prevent the incorrect information from participating into the decoding process and this shows that it is better not to use the wrong information. After redecoding without these wrong-RS-checked clusters, there can also exist other oligo clusters that do not satisfy above conditions. Therefore, we repeat the redecoding method several times iteratively.

\subsubsection{Iterative Decoding}
\label{subsubsec:IterativeDecoding}
Now, we introduce the proposed iterative soft decoding method shown in Algorithm~\ref{alg1}. For inputs, we use stitched oligo reads with the correct length (152nt) after the PEAR algorithm \cite{Zhang} and we only use oligo reads with \textcolor{black}{exactly the same seed information compared to the pre-determined seed values from the encoding process.} As a preprocessing step, we calculate LLRs for 1st and 2nd bits of each position for the payload parts. On the other hand, we need to use hard values for the RS part, and thus we determine the hard values of nucleotides from the maximum values of $\Pi_j P_i^j({\rm A}), \Pi_j P_i^j({\rm C}), \Pi_j P_i^j({\rm G})$, and $\Pi_j P_i^j({\rm T})$, where $P_i^j(b)$ is $P_i^{j,b^\prime}(b)$ for a basecall $b^\prime$ of the $i$th position of the $j$th stitched oligo read corresponding to the target oligo cluster and $b,b^\prime\in\{{\rm A,C,G,T}\}$. Also, in Step 4 of Algorithm~\ref{alg1}, we use $n_{\rm re} = 3$ for our redecoding count. In most cases, increasing the number of redecoding count $n_{\rm re}$ improves the decoding performance. \textcolor{black}{However, considering both the improvement of decoding performance and computational complexity,} we choose an adequate number $n_{\rm re} = 3$ \textcolor{black}{heuristically.} \textcolor{black}{This makes} \textcolor{black}{the maximum number of iterative decoding be 3 in the worst case, however, this parameter can be adjusted if we exploit the increased computational complexity.}

\begin{algorithm}[h]
\renewcommand{\thealgorithm}{1}
 \caption{Proposed iterative soft decoding algorithm.}
 \label{alg1}
 \begin{algorithmic}[1]
 \renewcommand{\algorithmicrequire}{\textbf{Input:} }
 \REQUIRE $L_m$ randomly sampled stitched oligo reads after the PEAR algorithm with \textcolor{black}{the pre-determined} seed values, the maximum re-decoding number $n_{\rm re}$, and LLRs and RS parity of those stitched oligo reads from a preprocessing step.
 \renewcommand{\algorithmicrequire}{\textbf{Output:} }
 \REQUIRE $16050 \times 256$ information bits.
\renewcommand{\algorithmicrequire}{\textbf{Initialization:} }
 \REQUIRE $i=0$.
\renewcommand{\algorithmicrequire}{\text{Step 1)} }
 \REQUIRE Perform soft decoding of $256$ binary LT codes (= $128$nt).
\renewcommand{\algorithmicrequire}{\text{Step 2)} }
 \REQUIRE For each decoded oligo, do RS decoding.
\renewcommand{\algorithmicrequire}{\text{Step 3)} }
 \REQUIRE If all RS decoding succeeds without errors in seed positions, find punctured information bits using decoded parity bits and skip the following steps.
\renewcommand{\algorithmicrequire}{\text{Step 4)} }
 \REQUIRE If $i=n_{\rm re}$, declare decoding failure and skip the following step.
\renewcommand{\algorithmicrequire}{\text{Step 5)} }
 \REQUIRE If $i<n_{\rm re}$ and oligos with RS decoding failure or seed position errors exist, remove those oligos and reconstruct $H$ matrix. Then, go to Step 1) with initial LLRs for the other bits of oligos and $i\leftarrow i+1$.
 \end{algorithmic}
 \end{algorithm}

\section{Result}
\label{sec:Result}
To show the superiority of the proposed decoding algorithm compared to the state-of-the-art algorithm, we \textcolor{black}{employ} two decoding algorithms \textcolor{black}{for two sequencing data}. One is the hard decoding method from the previous work \cite{Jeong} and the other is the proposed soft decoding method. For a fair comparison, we conduct these experiments with the assumption of knowing the seed information as we mentioned in \textcolor{black}{Section~\ref{subsec:ExperimentalDesign}} using $data A$ and $data B$.

\subsection{Experimental Result}
\label{subsec:ExperimentalResult}
\begin{figure}[!t]
    \includegraphics[scale=0.58]{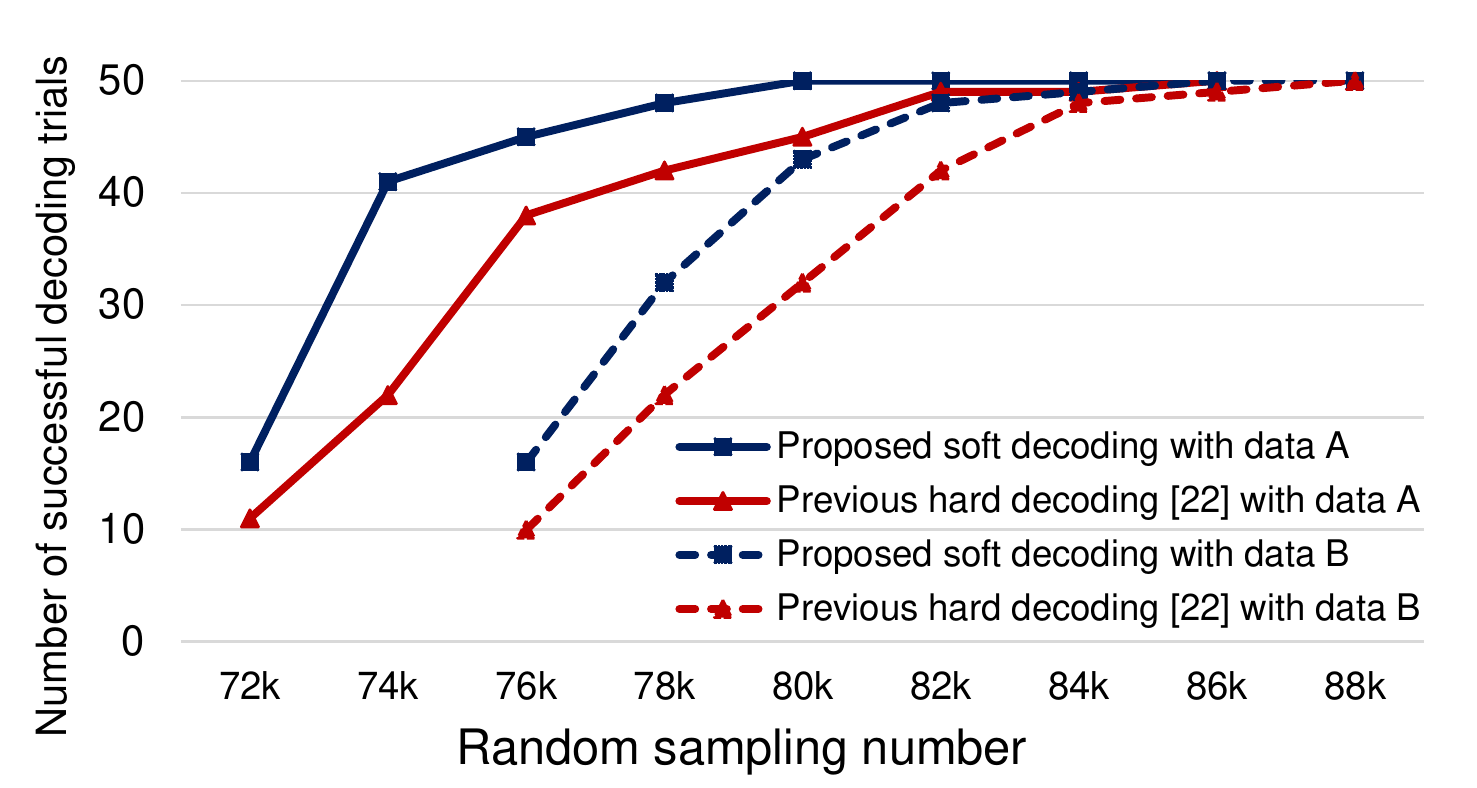}
    \caption{Decoding results of the proposed soft decoding and previous hard decoding algorithms in both $data A$ and $data B$ with 50 trials each.}
    \label{fig5}
\end{figure}

We randomly sample a small number of sequenced oligo reads first and add 2000 sequenced oligo reads for each sampling point to find a perfect recovery point. Note that $L_m$ in Algorithm~\ref{alg1} is smaller or equal to the random sampling numbers in Table~\ref{table1} due to the conditions of the exact oligo length (152nt) and exact seed position information. We conduct 50 trials for each random sampling and count the number of decoding \textcolor{black}{success} out of 50 trials. The result is summarized in Table~\ref{table1} and shown in Fig.~\ref{fig5}.

The result shows that the proposed method reaches the perfect recovery point at the lower random sampling number, implying a better decoding performance. Compared to the hard decoding method in \cite{Jeong}, we could reduce 2000$\sim$6000 sequenced oligo reads, which are 2.3\%$\sim$7.0\% decrease of reading \textcolor{black}{numbers. In fact, we make sure that a smaller number of sequenced oligo reads can imply a reduction of the reading cost in our experiment. Our experiments are conducted with the Illumina sequencing equipment, and one Illumina flow cell is always needed during each sequencing process. However, a smaller number of required oligo reads for decoding success} \textcolor{black}{means that we can use a smaller portion of the flow cell for reading our data.}

\begin{table}[!t]
\caption{\textcolor{black}{Experimental} results of the proposed soft decoding and previous hard decoding algorithms with 50 trials each}
\label{table1}
\begin{tabular}{lllll}
\hline
                                                           & \multicolumn{2}{c}{$data A$}                                                                                                                  & \multicolumn{2}{c}{$data B$}                                                                                                                  \\ \hline
\begin{tabular}[c]{@{}l@{}}Random\\ sampling\\ number\end{tabular} & \begin{tabular}[c]{@{}l@{}}Proposed  \\ soft\\ decoding\end{tabular} & \begin{tabular}[c]{@{}l@{}}Previous  \\ hard\\ decoding\cite{Jeong}\end{tabular} & \begin{tabular}[c]{@{}l@{}}Proposed  \\ soft\\ decoding\end{tabular} & \begin{tabular}[c]{@{}l@{}}Previous  \\ hard\\ decoding\cite{Jeong}\end{tabular} \\ \hline
72k                                                                & \textcolor{black}{16}                                                                   & 11                                                                   &                                                                      &                                                                      \\
74k                                                                & \textcolor{black}{41}                                                                   & 22                                                                   &                                                                      &                                                                      \\
76k                                                                & 45                                                                   & 38                                                                   & 16                                                                   & 10                                                                   \\
78k                                                                & 48                                                                   & 42                                                                   & 32                                                                   & 22                                                                   \\
80k                                                                & \textbf{50}                                            & 45                                                                   & 43                                                                   & 32                                                                   \\
82k                                                                & \textbf{50}                                            & 49                                                                   & 48                                                                   & 42                                                                   \\
84k                                                                & \textbf{50}                                            & 49                                                                   & 49                                                                   & 48                                                                   \\
86k                                                                & \textbf{50}                                            & \textbf{50}                                            & \textbf{50}                                            & 49                                                                   \\
88k                                                                &                                                                      &                                                                      &  \textbf{50}                                            & \textbf{50}                                            \\ \hline
\end{tabular}
\end{table}

\textcolor{black}{Also, our decoding algorithm is superior even though we do not introduce any other preprocessing methods during the clustering step.} If we add and modify other preprocessing methods as in \cite{Jeong} for the soft decoding method, such as Hamming-distance based clustering or Hamming-distance based discarding wrong sequenced oligo reads, we may have more improved decoding performance than the current results shown in Table~\ref{table1}. Note that we have total 18000 kinds of oligo sequences in the experiments and 2000$\sim$6000 sequenced oligo reads are a deep portion of the sequencing environments. Also, the proposed soft decoding algorithm outperforms the state-of-the-art algorithm for both of the two sequencing \textcolor{black}{data}, which implies the consistency of our algorithm.

\subsection{Analysis}
\label{subsec:Analysis}
The main reason why the decoding performance of $data A$ is better than $data B$ at the same random sampling point is that the ratios of perfectly stitched oligo reads after the alignment algorithm are different for $data A$ and $data B$ as shown in Table~\ref{table2}. We use the PEAR algorithm \cite{Zhang} for stitching the forward-reads and reverse-reads, and 83.83\% of the sequenced oligo reads in $data A$ are merged with the correct length of 152nt but 78.64\% of the sequenced oligo reads in $data B$ are merged with 152nt. Since we sample the sequenced oligo reads before the PEAR algorithm, $data A$ would have more stitched oligo reads after the PEAR algorithm than $data B$.

\begin{table}[!t]
\caption{Comparison of statistical properties of $data A$ and $data B$ for stitched oligo reads with the correct length (152nt)}
\label{table2}
\begin{tabular}{lllll}
\hline
Statistics                                                                                            & $data A$   & $data B$   \\ \hline
Correct-length (152nt) \\PEAR passing rate                                                         & 83.83\%   & 78.64\%    \\
\begin{tabular}[c]{@{}l@{}}Average number of \\position errors per base \end{tabular}           & $1.100 \times 10^{-3}$ & $9.604 \times 10^{-4}$ \\
\begin{tabular}[c]{@{}l@{}}Average number of \\insertion \& deletion errors per base \end{tabular} & $1.237 \times 10^{-5}$ & $1.744 \times 10^{-5}$ \\
\begin{tabular}[c]{@{}l@{}}Average number of \\substitution errors per base \end{tabular}       & $9.858 \times 10^{-4}$ & $8.352 \times 10^{-4}$ \\ \hline
\end{tabular}
\end{table}

\begin{table}[!t]
\caption{Effects of each decoding technique in the proposed soft decoding method in $data B$ \textcolor{black}{(LLR calculation and redecoding)}}
\label{table3}
\begin{tabular}{lllll}
\hline
\begin{tabular}[c]{@{}l@{}}LLR\\ calculation\end{tabular}          & \multicolumn{2}{l}{Chandak's method \cite{Chandak}}                                                                                           & \multicolumn{2}{l}{Proposed method}                                                                                            \\ \hline
\begin{tabular}[c]{@{}l@{}}Random\\ sampling\\ number\end{tabular} & \begin{tabular}[c]{@{}l@{}}without\\ redecoding\end{tabular} & \begin{tabular}[c]{@{}l@{}}with\\ redecoding\end{tabular} & \begin{tabular}[c]{@{}l@{}}without\\ redecoding\end{tabular} & \begin{tabular}[c]{@{}l@{}}with\\ redecoding\end{tabular} \\ \hline
76k                                                                & 0                                                               & 2                                                            & 0                                                               & \textbf{16}                                    \\
78k                                                                & 0                                                               & 21                                                           & 4                                                               & \textbf{32}                                    \\
80k                                                                & 9                                                               & 38                                                           & 10                                                              & \textbf{43}                                    \\
82k                                                                & 19                                                              & 46                                                           & 19                                                              & \textbf{48}                                    \\
84k                                                                & 23                                                              & 48                                                           & 21                                                              & \textbf{49}                                    \\
86k                                                                & 22                                                              & 48                                                           & 24                                                              & \textbf{50}                                    \\ \hline
\end{tabular}
\end{table}

For error statistics, we collect all the correct-length stitched oligo reads and compare them with our original data in Table~\ref{table2}. There exist differences between the error conditions for the two sequencing data and this would have effected the decoding performances as well. Also, this implies that for the same synthesized \textcolor{black}{pool} and the same sequencing \textcolor{black}{method}, the error statistics for each sequencing may be different a little. By the way, the error rates in both \textcolor{black}{sequenced data} are in an acceptable level of the Illumina sequencing environment.

To show the effects of each decoding technique in the proposed decoding algorithm, we carry out some more experiments with $data B$ in Table~\ref{table3}. For the proposed LLR calculation method, we compare it with Chandak's soft decoding method \cite{Chandak} based on the basecalling system. For the redecoding method, we conduct these two soft decoding methods in a way that incorporates and does not incorporate the redecoding system. Table~\ref{table3} shows that the proposed soft decoding method makes the best performance and each technique is necessary for our contributions. \textcolor{black}{Specifically, we} find that without the redecoding system, the channels of the DNA storage system are very unstable and we could even find that Chandak's method makes the performance reversal in some points. \textcolor{black}{For LLR calculation, it is noted that our method is superior in most cases. Therefore,} we can claim that the soft decoding method \textcolor{black}{using} soft information \textcolor{black}{(Q-scores extracted from the sequenced data and channel statistics)} is very useful for the DNA storage system. \textcolor{black}{Further, we note that} the redecoding method makes the bigger contribution here compared to the new formula of the LLR calculation\textcolor{black}{. Without} the redecoding system, our performance is even worse than the previous work \cite{Jeong} in Table~\ref{table1}.

\section{Discussion and Future Work}
\label{sec:DiscussionandFutureWork}
\textcolor{black}{In Section~\ref{sec:Result}, we consider two non-constrained data ($data A$ and $data B$) due to the condition that the decoder knows the pre-determined seed values. However, one may wonder whether the performance improvement of the proposed decoding algorithm is valid only for non-constrained data. Therefore, we employ our decoding algorithm using the experimental data of the constrained pool in \cite{Jeong} with 50 trials, only adding an assumption that we know the seed information of the constrained pool (Table~\ref{table4}). We can see that our soft decoding algorithm also performs better than the hard decoding algorithm in the constrained pool. This implies that our decoding algorithm is superior regardless of homopolymer-run length and GC-content constraints.}

\begin{table}[!t]
\caption{\textcolor{black}{Experimental result of the proposed soft decoding and previous hard decoding algorithms with 50 trials each from the data in the constrained pool of \cite{Jeong}}}
\label{table4}
\color{black}
\begin{tabular}{lll}
\hline
\multicolumn{3}{c}{Experimental data of the constrained pool in \cite{Jeong}}                                                                                             \\ \hline
Random sampling number & \begin{tabular}[c]{@{}l@{}}Proposed\\ soft decoding\end{tabular} & \begin{tabular}[c]{@{}l@{}}Previous\\ hard decoding \cite{Jeong}\end{tabular} \\ \hline
72k                    & 12                                                               & 10                                                                        \\
74k                    & 28                                                               & 22                                                                        \\
76k                    & 43                                                               & 37                                                                        \\
78k                    & 49                                                               & 43                                                                        \\
80k                    & 50                                                               & 49                                                                        \\
82k                    & 50                                                               & 50                                                                        \\ \hline
\end{tabular}
\end{table}

\begin{figure}[!t]
    \includegraphics[scale=0.42]{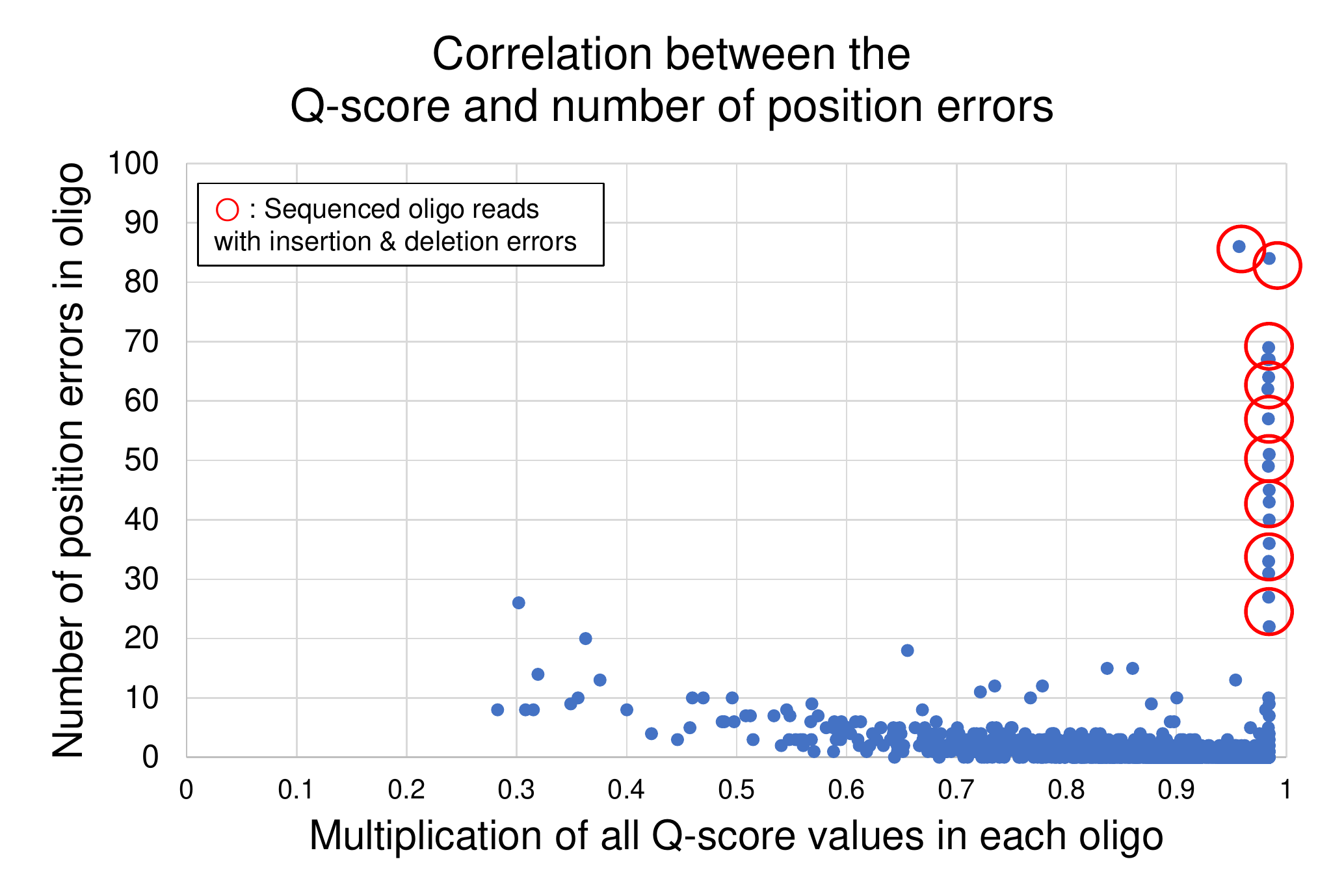}
    \caption{Correlation between the Q-score and number of position errors in each oligo cluster. This graph is from one of the 50 trials at the successful decoding point, 86k of $data B$.}
    \label{fig6}
\end{figure}

Besides the aforementioned algorithms, we have tried various attempts to improve the decoding performance. As we once used a multiplication of Q-score values for ordering the oligo cluster groups in the previous work \cite{Jeong}, we also try to use it in the soft decoding algorithm. The expression becomes:
\begin{equation}
P_j = \prod^{152}_{i=1}P_i^{j,b}(b) = \prod^{152}_{i=1}(1-10^{- \frac{Q_i^j(b)}{10}}),
\nonumber
\end{equation}
and we also calculate the correlation between this $P_j$ value and the number of position errors for each oligo cluster from the original encoded data in Fig.~\ref{fig6}. This time, we find out that some of the high Q-scored oligo reads with large errors (marked with red circles in Fig.~\ref{fig6}) are discarded during the redecoding process and many of them have insertion and deletion errors in them. It means that the proposed iterative soft decoding method can be effective for handling some insertion and deletion errors. Meanwhile, we try to discard some of the low Q-scored oligo reads during the decoding stage but the effect is relatively marginal (0 or 1 more decoding success out of 50 trials in Table~\ref{table1}) and it is effective in large random sampling points but sometimes gives worse performance in low random sampling points. However, since there exists the correlation between the multiplication of Q-scores and the number of errors, discarding some of the low Q-scored oligo reads may be effective. Thus, we believe that figuring out this correlation to the decoding performance is an important future work.

Moreover, when we determine the LLR calculation formula, we observe that providing LLR reliability differentially for each oligo cluster according to the size gives a better decoding performance for some random sampling points. Specifically, we assign the lowest LLR values for the clusters size of one, a little larger LLR values for the clusters size of two, and even more larger LLR values for the clusters size of three and more and this eventually improves our decoding performance a bit. We think this is because information in bigger size clusters is more trustful than information in smaller size clusters. We could not find any other theoretical clues and experimental criteria to determine the amount of variables to be multiplied to reduce the LLR values, leaving the differential distribution for LLR calculation for the future work.

In fact, soft decoding may require \textcolor{black}{the} more cost in the perspective of \textcolor{black}{the} computing power and computational complexity compared to the hard decoding method. However, synthesis and sequencing processes in DNA storage need much more resources and time and thus using a little more computing operations does not have a significant impact in terms of the overall procedure. By taking a little cost in computing power, we can save a lot of reading cost in DNA sequences and may even affect to increase the code rate of the entire data. Since the main target of DNA storage is cold data, we believe this assumption is reasonable.

Indeed, there are several ECCs that have better performance than the LT code for soft decoding. In the fountain code, a RaptorQ code \cite{Raptor} has been developed as an advanced code from the LT code. Also, an LDPC code is well known to have one of the best error correcting performances in soft decoding ~\cite{MacKay}. Our coding structure of the LT code and RS code in ~\cite{Science} is designed for the hard decoding method, not caring about the soft decoding method. Therefore, we are sure that if we conduct another experiment with well designed ECCs to fit with calculated soft information, our proposed algorithm would improve the performance dramatically compared to existing decoding methods. Further, the advantage of using LT codes is that we can control some constraints on the \textcolor{black}{homopolymer-run length} and GC-content. However, as shown in \cite{Jeong}, those constraints can also be managed through signal processing algorithms. Thus, other methods without using LT codes should be considered for DNA storage. The extreme case in the perspective of error correcting codes may be using ECCs with the best soft decoding performance dealing with position errors regardless of any other constraints.

\section{Conclusion}
\label{sec:Conclusion}
In this paper, we proposed a new iterative soft decoding algorithm for the DNA storage system, where soft information can be obtained from FASTQ files and channel statistics. By using conditional \textcolor{black}{base transition} rates and Q-scores, we deduced four probabilities of bases for each position of stitched oligo reads and suggested a new way of LLR calculation which can be used in the error correction and detection in the DNA sequencing area. From calculated LLR values, soft decoding of LT codes with the redecoding system was used in the proposed decoding algorithm. The proposed decoding algorithm outperformed the state-of-the-art algorithm in \cite{Jeong} and the existing soft decoding algorithm in \cite{Chandak} in terms of the required number of sequenced oligo reads, which means that the reading cost of the sequencing process can be reduced by our proposed decoding algorithm.




\begin{thebibliography}{00}

\bibitem{Church}
G. M. Church, Y. Gao, and S. Kosuri, "Next-generation digital information storage in DNA," {\em Science}, vol. 337, no. 6102, pp. 1628--1628, 2012.

\bibitem{Dong}
Y. Dong, F. Sun, Z. Ping, Q. Ouyang, and L. Qian, "DNA storage: research landscape and future prospects," {\em National Science Review}, vol. 7, no. 6, pp. 1092--1107, 2020.

\bibitem{Cock}
P. J. A. Cock, C. J. Fields, N. Goto, M. L. Heuer, and P. M. Rice, "The Sanger FASTQ file format for sequences with quality scores, and the Solexa/Illumina FASTQ variants," {\em Nucleic Acids Research}, vol. 38, no. 6, pp. 1767--1771, 2010.

\bibitem{Xu}
C. Xu, C. Zhao, B. Ma, and H. Liu, "Uncertainties in synthetic DNA-based data storage," {\em Nucleic Acids Research}, vol. 49, no. 10, pp. 5451--5469, 2021.

\bibitem{El-Shaikh}
A. El-Shaikh, M. Welzel, D. Heider, and B. Seeger, "High-scale random access on DNA storage systems," {\em NAR Genomics and Bioinformatics}, vol. 4, no. 1, 2022, Art. no. lqab126.

\bibitem{Ross}
M. G. Ross, C. Russ, M. Costello, A. Hollinger, N. J. Lennon, R. Hegarty, C. Nusbaum, and D. B. Jaffe, "Characterizing and measuring bias in sequence data," {\em Genome Biology}, vol. 14, no. 5, pp. 1--20, 2013.

\bibitem{Schwartz}
J. J. Schwartz, C. Lee, and J. Shendure, "Accurate gene synthesis with tag-directed retrieval of sequence-verified DNA molecules," {\em Nature Methods}, vol. 9, no. 9, pp. 913--915, 2012.

\bibitem{Cao_nano}
B. Cao, X. Zhang, J. Wu, B. Wang, Q. Zhang, and X. Wei, "Minimum free energy coding for DNA storage," {\em IEEE Transactions on NanoBioscience}, vol. 20, no. 2, pp. 212--222, 2021.

\bibitem{Cao_bio}
B. Cao, X. Li, X. Zhang, B. Wang, Q. Zhang, and X. Wei, "Designing uncorrelated address constrain for DNA storage by DMVO algorithm," {\em IEEE/ACM Transactions on Computational Biology and Bioinformatics}, vol. 19, no. 2, pp. 866--877, 2022.

\bibitem{Wu}
J. Wu, Y. Zheng, B. Wang, and Q. Zhang, "Enhancing physical and thermodynamic properties of DNA storage sets with end-constraint," {\em IEEE Transactions on NanoBioscience}, vol. 21, no. 2, pp. 184--193, 2022.

\bibitem{Yin}
Q. Yin, Y. Zheng, B. Wang, and Q. Zhang, "Design of constraint coding sets for archive DNA storage", {\em IEEE/ACM Transactions on Computational Biology and Bioinformatics}, \textcolor{black}{vol. 19, no. 6, pp. 3384--3394, 2021.}

\bibitem{Goldman}
N. Goldman, P. Bertone, S. Chen, C. Dessimoz, E. M. LeProust, B. Sipos, and E. Birney, "Towards practical, high-capacity, low-maintenance information storage in synthesized DNA," {\em Nature}, vol. 494, no. 7435, pp. 77–-80, 2013.

\bibitem{Grass}
R. N. Grass, R. Heckel, M. Puddu, D. Paunescu, and W. J. Stark, "Robust chemical preservation of digital information on DNA in silica with error-correcting codes," {\em Angewandte Chemie International Edition}, vol. 54, no. 8, pp. 2552–-2555, 2015.

\bibitem{Bornholt}
J. Bornholt, R. Lopez, D. M. Carmean, L. Ceze, G. Seelig, and K. Strauss, "A DNA-based archival storage system," in {\em Proceedings of the Twenty-First International Conference on Architectural Support for Programming Languages and Operating Systems (ASPLOS)}, 2016, pp. 637-–649.

\bibitem{RandomAccess}
L. Organick, S. D. Ang, Y.-J. Chen, R. Lopez, S. Yekhanin, K. Makarychev, M. Z. Racz, G. Kamath, P. Gopalan, B. Nguyen, C. N. Takahashi, S. Newman, H.-Y. Parker, C. Rashtchian, K. Stewart, G. Gupta, R. Carlson, J. Mulligan, D. Carmean, G. Seelig, L. Ceze, and K. Strauss, "Random access in large-scale DNA data storage," {\em Nature Biotechnology}, vol. 36, no. 3, pp. 242--248, 2018.

\bibitem{Portable}
S. M. H. T. Yazdi, R. Gabrys, and O. Milenkovic, "Portable and error-free DNA-based data storage," {\em Scientific Reports}, vol. 7, no. 1, pp. 1--6, 2017.

\bibitem{Science}
Y. Erlich and D. Zielinski, "DNA Fountain enables a robust and efficient storage architecture," {\em Science}, vol. 355, no. 6328, pp. 950--954, 2017.

\bibitem{Chandak}
S. Chandak, K. Tatwawadi, B. Lau, J. Mardia, M. Kubit, J. Neu, P. Griffin, M. Wootters, T. Weissman, and H. Ji, "Improved read/write cost tradeoff in DNA-based data storage using LDPC codes," in {\em 2019 57th Annual Allerton Conference on Communication, Control, and Computing (Allerton)}, IEEE, 2019, pp. 147--156.

\bibitem{Hedges}
W. H. Press, J. A. Hawkins, S. K. Jones. Jr, J. M. Schaub, and I. J. Finkelstein, "HEDGES error-correcting code for DNA storage corrects indels and allows sequence constraints," in {\em Proceedings of the National Academy of Sciences}, vol. 117, no. 31, pp. 18489--18496, 2020.

\bibitem{Wang}
Y. Wang, M. Noor-A-Rahim, J. Zhang, E. Gunawan, Y. L. Guan, and C. L. Poh, "Oligo design with single primer binding site for high capacity DNA-based data storage," {\em IEEE/ACM Transactions on Computational Biology and Bioinformatics}, vol. 17, no. 6, pp. 2176--2182, 2019.

\bibitem{Cao_npj}
B. Cao, X. Zhang, S. Cui, and Q. Zhang, "Adaptive coding for DNA storage with high storage density and low coverage," {\em NPJ Systems Biology and Applications}, vol. 8, no. 1, pp. 1--12, 2022.

\bibitem{Jeong}
J. Jeong, S.-J. Park, J.-W. Kim, J.-S. No, H. H. Jeon, J. W. Lee, A. No, S. Kim, and H. Park, "Cooperative sequence clustering and decoding for DNA storage system with fountain codes," {\em Bioinformatics}, vol. 37, no. 19, pp. 3136--3143, 2021.

\bibitem{Kang}
S. Kang, Y. Gao, J. Jeong, S.-J. Park, J.-W. Kim, J.-S. No, H. Jeon, J. W. Lee, S. Kim, H. Park, and A. No, "Generative Adversarial Networks for DNA Storage Channel Simulator," \textcolor{black}{{\em IEEE Access}, vol. 11, pp. 3781--3793, 2023.}

\bibitem{Luby}
M. Luby, "LT codes," in {\em Proceedings of the 43rd Annual IEEE Symposium on Foundations of Computer Science, 2002}, IEEE Computer Society, pp. 271--271, 2002.

\bibitem{Han}
Y. Han and W. E. Ryan, "Low-floor decoders for LDPC codes," {\em IEEE Transactions on Communications}, vol. 57, no. 6, pp. 1663--1673, 2009.

\bibitem{Palanki}
R. Palanki and J. S. Yedidia, "Rateless codes on noisy channels," in {\em Proceedings of IEEE International Symposium on Information Theory (ISIT), 2004}, IEEE, pp. 37, 2004.

\bibitem{Zhang}
J. Zhang, K. Kobert, T. Flouri, and A. Stamatakis, "PEAR: a fast and accurate Illumina Paired-End reAd mergeR," {\em Bioinformatics}, vol. 30, no. 5, pp. 614--620, 2014.

\bibitem{Illumina}
Illumina Inc., "bcl2fastq Conversion User Guide," Illumina Inc., San Diego, CA, USA. Version 1.8.4. pp. 23--24.

\bibitem{Lu}
X. Lu, J. Jeong, J.-W. Kim, J.-S. No, H. Park, A. No, and S. Kim, "Error rate-based log-likelihood ratio processing for low-density parity-check codes in DNA storage," {\em IEEE Access}, vol. 8, pp. 162892--162902, 2020.

\bibitem{Raptor}
A. Shokrollahi, "Raptor codes," {\em IEEE Transactions on Information Theory}, vol. 52, no. 6, pp. 2551--2567, 2006.

\bibitem{MacKay}
D. J. C. MacKay and R. M. Neal, "Near Shannon limit performance of low density parity check codes," {\em Electronics Letters}, vol. 33, no. 6, pp. 457--458, 1997.

\end{thebibliography}
\end{document}